# Industrial Strength Software in Computer Based Engineering Education (CBEE): a Case Study

*M. V. Bogdanov[1], D. Kh. Ofengeim[1], A.V. Kulik[1], D.V. Zimina[1], M.S. Ramm[2], A.I. Zhmakin[1,3]*

[1] Soft-Impact, Ltd., P.O.Box 83, 27 Engels av., St.Petersburg, 194156 Russia

[2] Semiconductor Technology Research, Inc., P.O.Box 70604, Richmond, VA, USA

[3] A.F. Ioffe Physical Technical Institute, Russian Academy of Sciences, St.Petersburg, 194121 Russia

**Abstract:** Challenging problems of modern engineering education and a role of information technology are reviewed. An importance of simulation of real world problems in both graduate and lifelong/corporate education is discussed. It is proposed to integrate the hypermedia theory courseware with industrial strength simulation software. As a proof of the concept an educational environment "Heat and Mass Transfer in Advanced Semiconductor Technology" has been developed.

**Interactive Demonstration:** A demo version of the Integrated Educational Environment (IEE) "Heat and Mass Transfer in Advanced Semiconductor Technology" can be downloaded from
http://www.semitech.us/products/IEE.

# 1 Introduction

In today's rapidly changing knowledge-based economy, learning is needed to survive. This is true for individuals, for organizations, for communities, for nations. However, numerous researches show that the traditional didactic lecture practice supports memorization of factual information but is by far less effective in promoting the ability to apply the learned concepts to the real world problems (McCray, DeHaan and Schuck, 2003). Too often graduates "*know everything, but can do nothing*" (Eyerer, Hefer and Krause, 2000). Modern education should fully exploit the possibilities provided by advances in information technologies (IT).

These concerns are widely recognized at the different levels – from faculty to government. E.g., the strategic goals for US education listed in Report "Education the engineer of 2020: adapting engineering education to the new century" (NAE, 2005) include the following ones: "*how to enrich and broaden engineering education so that graduated will be better prepared to work in a constantly changing global economy; institutions must teach students to be lifelong learners; growing need for interdisciplinary and system-based approaches*". Similar programs are considered by other nations. The Korean government announced the Basic Framework for National Human Resource Development aiming at the creation of an educational information infrastructure and a cyber education system [1]. The European Commission has adopted a Communication on "Making a European Area of Lifelong Learning a Reality" [2]. Probably, the most urgent need for novel approaches to education is in Science, Technology, Engineering, and Mathematics (STEM).

The aim of the paper is to review the problems of modern engineering education and to discuss how most of them could be, to a great extent, resolved by using next-generation Computer-Based Education software environment - Integrated Educational Environment (IEE). The paper is organized as follow. Main requirements to outcome of engineering education and aspects of the learning process are discussed in the next section. The role of information technologies in education is analyzed in section

3. Design principles and basic features of IEE for Advanced Semiconductor Technology are described in section 4. Closing remarks are contained in Conclusions.

## 2   Modern engineering education

Engineers "*scope, generate, evaluate, and realize ideas*" (Sheppard, 2003). Improvement of engineering education is, first of all, a better alignment of engineering curricula with challenges graduates will face in the workplace (NAE, 2005). Extended list of requirements to outcome of education formulated by the Accreditation Board for Engineering and Technology (ABET) includes, among others, the following items (Prados, Peterson and Lattuca, 2005) : *"(a) an ability to apply knowledge of mathematics, science, and engineering; ... (j) a knowledge of contemporary issues; (k) an ability to use the techniques, skills, and modern engineering tools necessary for engineering practice."* Education evolution should proceed along the well-known Bloom's taxonomy of the extent of knowledge mastering (Jensen and Wood, 2000): Knowledge (list or recite); Comprehension (explain or paraphrase); Application (calculate, solve, determine or apply);  Analysis (compare, contrast, classify, categorize, derive, model); Synthesis (create, invent, predict, construct, design, imagine, improve, produce, propose); Evaluation  (judge, select, decide, critique, justify, verify, debate, assess, recommend). Zhurakovsky, Pokholkov and Agranovich (2001) stress that the student in high technologies needs to go out of the area of knowledge into the area of practice. Modern requirements for undergraduate and corporate education in engineering are similar: as  Bourne, Harris and Mayadas  (2005) note, a distinction between education and training is being blurred: "*In engineering, yesterday's education can become today's training overnight*".

Success of education depends on both teaching practices and student abilities/preferences. The great pedagogical challenge has been formulated almost two decades ago by Felder and Silverman (1988): "*What can be done to reach students whose learning styles are not addressed by standard methods of engineering education?*"

### 2.1 Learning styles

Students have different levels of motivation and different responses to specific instructional practices. name three categories of diversity that have important implications for teaching and learning: differences in students' learning styles, approaches to learning and intellectual development levels. Learning styles are "characteristic cognitive, affective, and psychological behaviors that serve as relatively stable indicators of how learners perceive, interact with, and respond to the learning environment" (Keef, 1979). Students' approach to study could be characterized in one of three ways (Entwistle, 1988): 1) a reproducing orientation leading to a surface approach, relying on memorization; 2) a meaning orientation resulting in a deep approach, exploring the limits of applicability of new material; 3) an achieving orientation ending in a strategic approach, doing whatever is necessary to get the highest grade.

Correlation of learning styles with student's attributes has been studied in numerous works. A number of classification systems have been proposed, the most widely used being, probably, The Myers-Briggs Type Indicator (MBTI), Kolb's Learning Style Model, Herrmann Brain Dominance Instrument (HBDI), Felder-Silverman Learning Style Model  (Felder and Silverman, 1988, Felder, 1996, Jensen and Wood, 2000). The conclusion to be drawn by assessment of traditional engineering education is, however, the same whatever model one is using: the teaching patterns do not suit the majority of students. Traditional approach (by presenting lectures and requiring individual assignments) is oriented toward introverts, intuitors, thinkers, and judgers in terms of MBTI, Type 2 learners (Kolb model), strongly Quadrant A dominant (HBDI), intuitive, verbal, deductive, reflective, and sequential learners (Felder-Silverman model). To cope with the problem, approaches for "teaching to all types" (Felder, 1996) have been proposed as well as test-based identification of the student's preferences (Jensen and Wood, 2000).  National Academy of Engineering realizes the severity of this problem indicating as one of priority strategies better alignment of faculty skill sets with those needed to deliver the desired curriculum in light of the different learning styles of students (NAE, 2005).

The problem of gender and cultural style preferences is also actual. E.g., NAE states that engineering in US is a profession where minorities and women remains underrepresented (NAE, 2005). One of the recommendations of Workshop "Information Technology (IT)-Based Educational Materials" held in 2003 requires the creation of models on how people learn STEM concepts, including how retention rates can be improved for women and students from underrepresented groups [3]. Studies of minorities and female learning style dictate flexibility in teaching practices and patterns. E.g., it has been found that Hispanic students prefer conformity, peer-oriented learning and a high degree of structure while African American students tend to prefer inferential reasoning, to focus on people rather than things, and to be more proficient in nonverbal communications; when dealing with Native American students, teachers should minimize lecturing, place less emphasis on competition, and minimize teacher directions (Heredia, 1999). Women students respond better to the use of visualization and computers; female-friendly pedagogy includes active, cooperative learning (Ribando, Richards and O'Leary, 2004). USA national scale effort to improve the retention of under-represented engineers has been reported by Agogino and His (1995).

## 2.2 Teaching practices

There is a number of teaching practices, most common ones being listed by Spalter and Simpson (2000): laboratory, visualization, simulation, lecture and demonstration, case-study, role- playing, mastery learning, creative project, student teaching, playground, drill, behavior modification, and incidental learning. Similar but different taxonomy of "learning processes" has been given by Lytras, Pouloudi and Poulymenakou (2002): presentation, explanation, relation synthesis, analysis, evaluation, reasoning, problem solving, collaboration, learning story preparation. Only few of these approaches are used, as a rule, in teaching practice simultaneously. Nichols (2002) summarized best practice principles for XXI century education identified by a number of educators.

From engineering perspective, project-based learning (PBL) and case study (Hills and Tedford, 2003) are ones of the most promising practices. The acronym PBL is also used in the education literature to signify problem-based learning, in which abstract theoretical material is introduced in more familiar, everyday problem situations. The two PBL's have some common features, but they are nonetheless distinct pedagogical styles (Dym, Agogino, Eris, Frey and Leifer, 2005). Project-based learning provides students the possibility to "learn by doing" which differ greatly from traditional sequential "chalk and talk" approach (Delaney, Mitchell and Delaney, 2003). The introduction of PBL into educational courses is supported by the constructivist theory of learning, the most important principles of the latter being formulated by Kolari and Savander-Ranne (2003) as *"learning is a process; knowledge cannot be directly transmitted from the teacher to the learner, it will have to be constructed or reconstructed by the learner him/herself; goal-directed learning is a skill that can be learned"*.

To conclude this section, the main requirements to engineering education can be summarized as
- **practice orientation** (knowledge as a skill, not as a matter)**;**
- **multidisciplinarity and system-level approach;**
- **adaptativity to the learner's objectives, background, style and needs;**
- **lifelong learning** ("shelf-life" of knowledge in today's world is short).

Evidently, the system of engineering education needs a radical reformation to meet these challenges. Information Technology (IT) development provides the necessary means.

## 3 Computer-based education (CBE)

Unfortunately, IT enhanced education in all disciplines including engineering is almost exclusively understood as online learning, usually enriched by multimedia features. The benefits of online education have been listed in numerous papers (see, e.g. Evans and Haase, 2001). Its scale is impressive – over 2 million learners today; 50-75 % of more than 2,000 corporate universities are Web-based (Bourne et al., 2005). There is no consensus, however, on the quality of online education. According to some reports (Bourne et al., 2005, Aragon, Johnson and Shaik, 2002), there are no significant differences in learning outcomes for online and on-campus (face-to-face) students as measured by test scores while there are also claims referring to higher drop-out and poorer performance that *"e-learning is virtually guaranteed to fail"* (Van Liew, 2005). Thus, as Bartley and Golek (2004) note, at present there is no conclusive research concerning the effectiveness of online education.

Different terms are used for computer-based and online education. Recent analysis of taxonomy of IT educational technologies such as computer-based learning, distance learning, e-learning, online learning etc., mainly based on the technical aspects of the knowledge deliver rather than on the content improvement and enhancement, has been reported by Anohina (2005). This focus on technology in IT education reflects the common attitude that is righteously termed as "misunderstanding" [4] and "misconception" (Bourne et al., 2005) : "*Without appropriate, interactive and stimulating course content, the technology is useless*". Van Dam, Becker and Simpson (2005) state: "*Although IT had certainly some impact, it has become a cliché to note that education is the last field to take systematic advantage of IT*". Among the reasons listed by the authors, the most important ones seem to be the following two:

- The general *conservatism* of educational institutions;
- Inadequate investment in the creation of new dynamic and interactive **content.**

It seems that IT education terminology and its usage are somewhat confusing. Online (distance) education and computer-based education (CBE) are not synonyms. Distance learning has existed since the 19th century, when farmers used correspondence courses to bone up on plowing techniques. Even optimal use computer-based and Internet-based technology for delivery (Kulacki, Sakamoto and Swope, 2002), does not make, in our opinion, this kind of education "computer-based" one. Such transformation of traditional course by means of digital and telecommunication technologies does not fully exploit the possibility to enhanced education. As noted by Bartley and Golek (2004), "*A dull, unrewarding course taught in the classroom environment becomes even duller and less rewarding in the online environment*", be it recorded in a radio studio (Kulacki et al., 2002) or delivered live using video-conferencing software (Bhavnani, Bar-Cohen and Joshi, 1999). As Felder (2005a) stresses, "*using technology simply to archive traditional lectures or Power Point™ shows on the Web represents **a regression** rather than an improvement over traditional methods*". Thus, it is suggested to use "CBE' term when computer is used to enhance the learning material, not just to deliver it: "*Content is king"* (Gates, 1996).

There are unique features that (potentially) provide CBE advantage over traditional teaching approaches: **hypermedia** and, in case of engineering**, simulations.**

From an instructional perspective, a critical feature of hypermedia is that it provides a non-sequential information presentation that differs markedly from the text-based material used in conventional instructional systems (Gomes, Choy, Barton and Romagnoli, 2000). Thus, hypermedia based courseware can offer distributed, interactive, both synchronous and asynchronous, student-centred learning. Educational Adaptive Hypermedia Applications (EAHA) provide personalized views on the learning content and adaptive sequencing (navigation) over the learning content (Retalis and Papasalouros, 2005). A statistically significant increase in academic achievement has been recorded in comparison of hypermedia assisted and conventional learning (Zywno and Waalen, 2002).

While hypermedia is practically a standard in IT education ("*Content modules include digitally encoded lessons on specific topics, assembled textbooks, and interactive displays of information based*

*on inputs from users*" [3]), simulation as an educational tool is present in a lesser degree. The term 'simulation' is being applied in an increasingly broad manner and is sometimes used synonymously with 'animation' (Thomas and Milligan). For the purposes of this work a simulation is defined as having the following two key features:
1. There is a computer model of a real system that contains information on how the system behaves.
2. Experimentation can take place, i.e. changing the input to the model affects the output.

As a numerical model of a system, presented for a learner to manipulate and explore, simulations can provide a rich learning experience for a student and can be used in a two ways (Feisel and Rosa, 2005):

- as a pre-lab experience to give students (both undergraduate and corporate) some idea of what they will encounter in an actual experiment.

- as stand-alone substitutes for physical laboratory exercises, which is imperative for systems that are too large, too expensive, or too dangerous for physical measurements by undergraduate students.

E.g., only large universities usually offer semiconductor design courses because they have the funding available to setup and maintain a laboratory. Using simulation, however, gives the opportunity for smaller schools with a tighter budget to become more competitive by teaching these topics using the computer as a substitute for a semiconductor processing laboratory (Johnson and Ula, 1996).

Use of Java Applets in education is common. Their main advantage is zero overhead involved, such as in learning to use the courseware (Wie, 1998). The problems solved, however, are if not toy ones, but quite simple. Exploiting industrial strength software allows to introduce complex real-world problem solving into the classroom on a routine basis and thus to enhance carrier opportunities for students by giving them advanced experience in future work (Li and Liu, 2003; Mcintyre and Venkitachalam, 2005). Students need *"a tool for doing science and engineering, not a 'sandbox'"* [3].
The most sophisticated simulation tool used in a number of university courses is probably combined research and educational software environment "Interactive Ground Water (IGW)" for unified deterministic and stochastic groundwater modeling (Li and Liu, 2003) that won 2002 NSF Award as the best courseware. Other scarce examples include fluid dynamics and heat transfer software (Pieritz, Mendes, da Silva and Maliska, 2004; Ribando et al. 2004), acoustics and signal processing (Rahkila and Karjalainen, 1998), semiconductor processing (Johnson and Ula, 1996).

In summary, CBE certainly has a potential to meet educational challenges and to cope with learning styles problem providing tools that are "*pedagogically neutral*" (Kaw and Besterfield, 2004) and can be tuned for teacher/learner preferences.

Two pillars of Computer-Based Engineering Education (CBEE) are hypermedia and simulation. What was lacking so far is their two-way coupling that provides unprecedented flexibility in choosing learning and teaching patterns keeping in sight future real-world problems that will encounter the learner.

An example of such novel educational environment that combines theory course (in hypermedia form) and industrial strength simulation is described in the next section.

## 4  Integrated Educational Environment (IEE) "Heat and Mass Transfer in Advanced Semiconductor Technology"

Crystal growth seems to be a good example to illustrate the proposed approach. Firstly, the growth of crystals involves a number of physical processes: heat transfer, including conduction, convection and radiation (Figure 1), mass transfer, phase change, homogeneous and heterogeneous chemical reactions, electromagnetic phenomena (Bogdanov, Demina, Karpov, Kulik, Ramm and Makarov, 2003; Bogdanov, Ofengeim and Zhmakin, 2004). Therefore, one is forced to deal with the conjugated multidisciplinary problem. Moreover, since the crystal quality is of primary interest, the thermal stress, formation of point defects and dislocations in the crystal and their evolution during the growth and the post-growth processing should be also considered (Zhmakin, Kulik, Karpov, Demina, Ramm and Makarov, 2000). Secondly, growth equipment is expensive and only a few universities can afford it. Finally, growth process is long and complex, thus the price of operator's error is high in terms of time and money. Evidently, numerical simulation could be beneficial for both under/postgraduate education and corporate training.

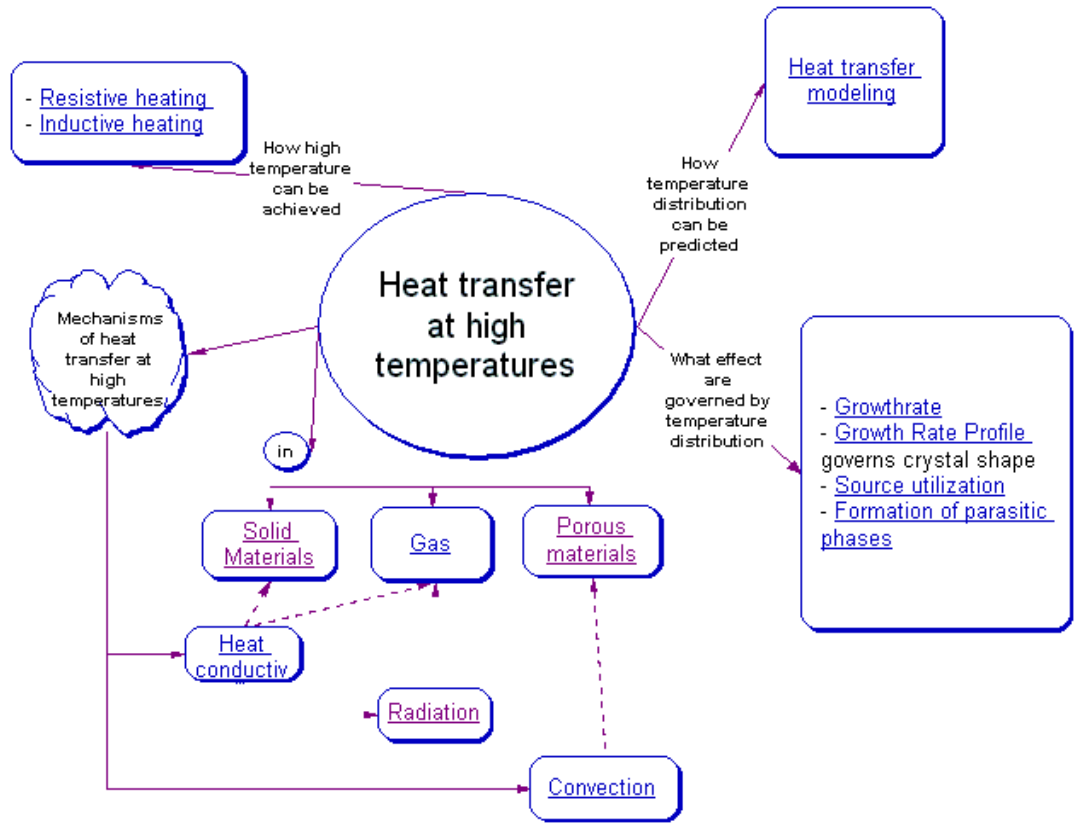

Figure 1: Heat transfer in crystal growth: a concept map

The drawback of using general-purpose commercial software is a long learning curve. A week or two (and frequently a lot longer) needed to master software may be acceptable in industry, while in the university class time is too valuable. As was noted by Rahkila and Karjalainen (1998), educational software should teach its content, not its usage. Moreover, as a rule, it is a time-consuming task to adjust general software to solution of multidisciplinary coupled problems. That's the reason that sometimes in-house software is developed for use in education (Ribando et al, 2004). On the other hand, research software usually differs in many aspects from industrial codes (Bogdanov et al., 2004) and thus does not allow the student to get familiar with modern engineering tools.

Probably, specialized codes for crystal growth are the best candidates for incorporation into educational software. Being designed and developed from a scratch as dedicated simulators for a limited class of growth processes, they could be made both powerful and user-friendly. It has been claimed that terms such "user-friendly" or "easy-to-learn" are ambiguous because they are subjective and thus unverifiable (Hooks, 1993). However, they can be measured in the relative units - one can easily compare two codes using the time needed to master the code operation by an uninitiated user or the time required for the specification of the geometry and the problem parameters.

IEE "Heat and Mass Transfer in Advanced Semiconductor Technology" has been developed using the software tool Virtual Reactor (ViR) [5] as the simulation engine. ViR employs advanced models of surface kinetics important for adequate predictions of the growth process. ViR has an easy-to-learn interface that allows the user either to describe the geometry manually or to import a CAD file. All geometric and process parameters entered by the user are checked automatically to belong to the corresponding interval of admissible values. The code has a number of features aimed at minimizing the user's efforts such as

- automatic initial identification of the blocks from the wire frame geometry
- automatic identification of the types of the inner boundaries
- one-time specification of the whole growth process
- automatic updating of the boundaries of the crystal and deposit blocks
- automatic processing of the topological changes of the computational domain (formation of new blocks and boundaries)
- automatic (re)generation of the unstructured grid in the modified and new blocks

Thus, IEE combines two components: 1) hypermedia textbook "Heat and Mass Transfer in Advanced Semiconductor Technology" (Figure 2) with links to laboratory tasks and 2) an adapted version of the Virtual Reactor software.

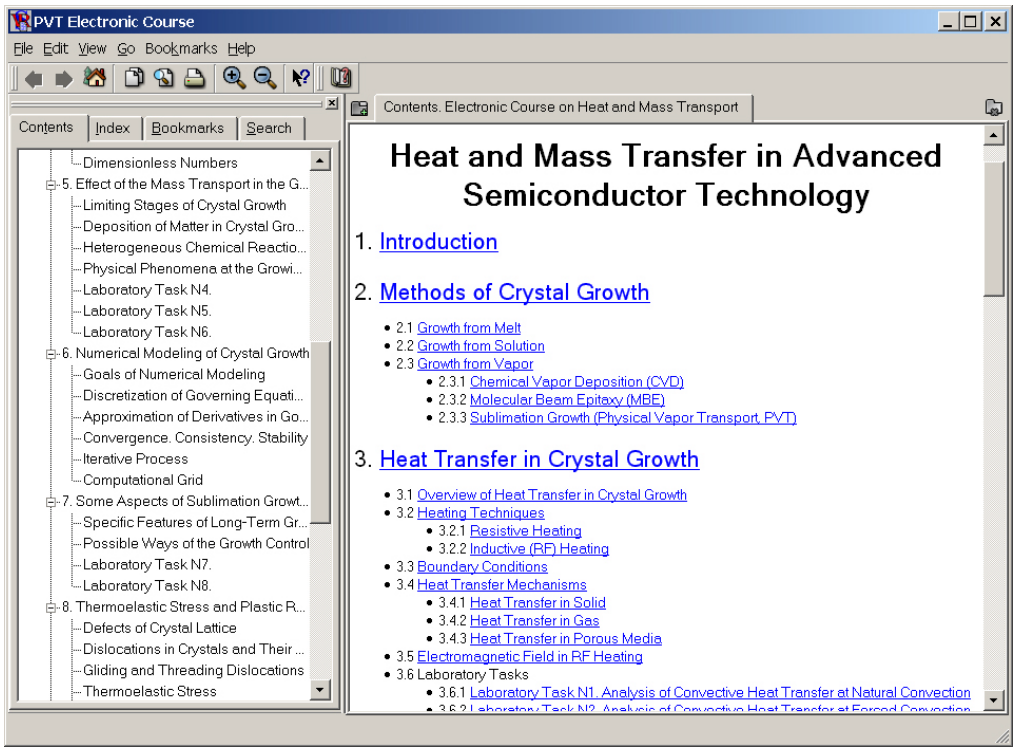

Figure 2: A snapshot of the textbook contents

A *Template* file specified independently for each laboratory task includes the following items:

- Reference to a pre-defined project file (an example of such file opened in the *Geometry* window is shown in Figure 3)

- Reference to a relevant Quick Help file (Figure 4) simplifying the accomplishment of the given task.

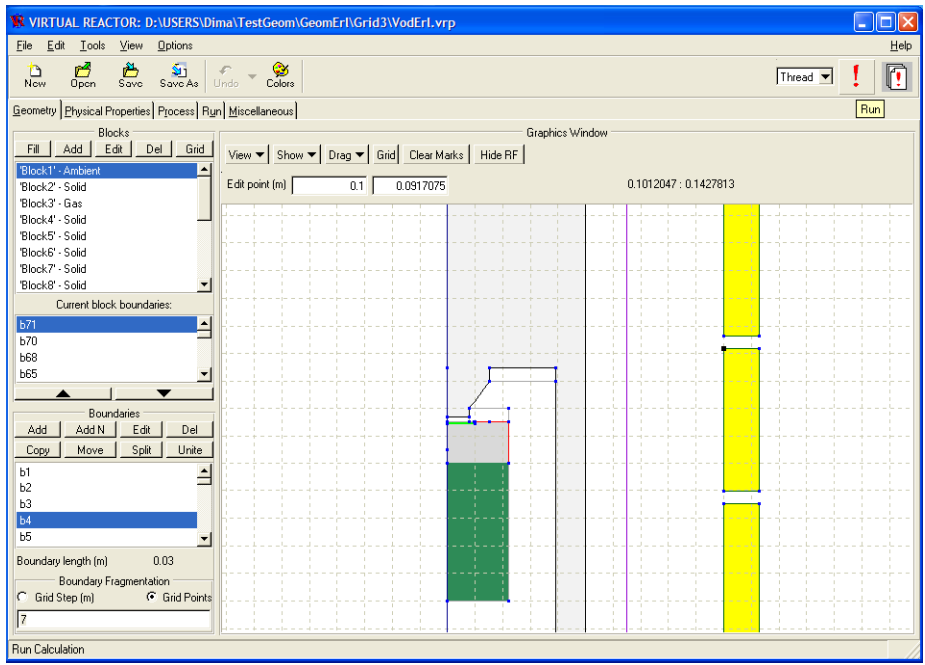

Figure 3: An example of the *Geometry w*indow

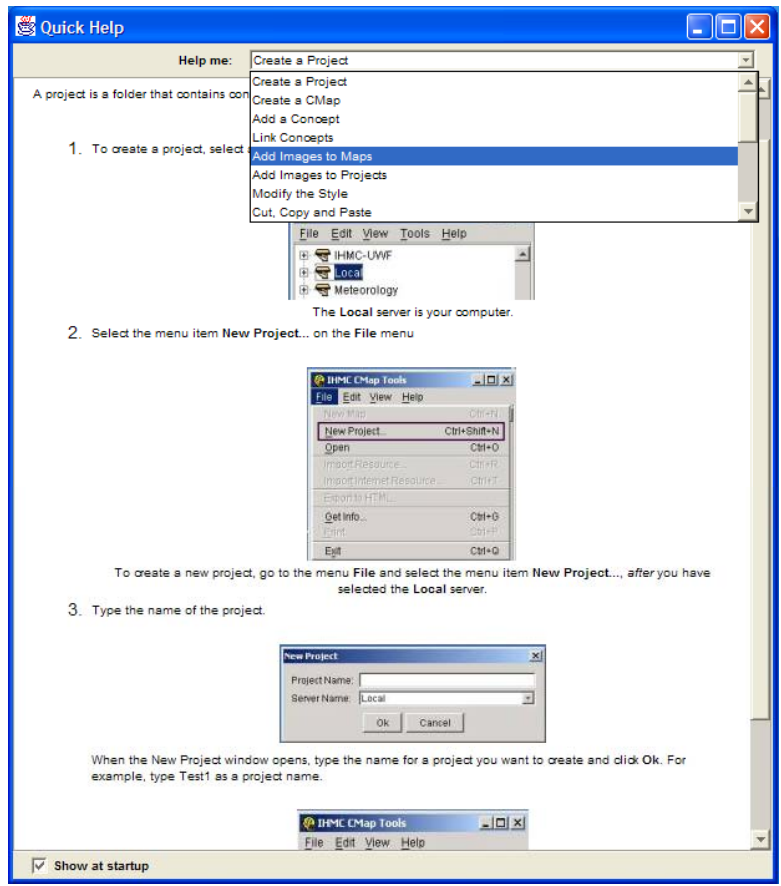

Figure 4: An example of the *Quick Help* window.

The simulation code incorporated in IEE code provides students and corporate engineers with exhaustive information about numerous physical processes responsible for the growth of bulk crystal and its quality. It provides an accurate solution of all major physical-chemical phenomena relevant to this method such as resistive or RF heating; conductive, convective and radiative heat transfer; mass transfer in gas and porous media; heterogeneous chemical reaction at catalytic walls and on the surface of powder granules; polycrystal formation; development of elastic strain and dislocations in the growing crystal; evolution of crystal and deposit shape, including partial facetting of the crystal surface. The problem is solved using a quasi-stationary formulation.

Unstructured grid is generated block-wise using Delauney algorithm, an advancing front method or their combination. Non-matched grids in the neighbour blocks are allowed. At the each virtual time step a number of subproblems is solved subsequently. Joule heat source distribution is determined by solution of Maxwell equations in the frequency domain. Global heat transfer analysis includes radiative transfer using the configuration factors. Darcy law is used to relate the velocity and the pressure distribution in the powder. Conjugate mass transfer in the gas and the powder using Hertz-Knudsen fluxes in the formulation of the boundary conditions for the species concentrations gives the growth rate at the catalytic surfaces. The type of the surface and local concentration of gas phase components determine the kind of growing crystal. Computation of thermal stresses and dislocation density is implemented as a post-processing procedure.

Transfer to the next global time instant includes the propagation of the crystal and deposit(s) boundaries, identification of new blocks and boundaries (if needed), the movement of the inductor (if specified by the user) and generation of unstructured or hybrid structured/unstructured grid in the new and altered blocks. A special optimization procedure for the growing front advancement has been developed that eliminate the effect of the numerical noise in the growth rate distribution and allows a stable evolution of the crystal and deposit(s) shape and a monitoring of the topological changes in the computational domain.

The straightforward use of the model is referred to as a *direct* problem. From a practical point of view the reversed formulation is more useful: how one should change the equipment design or the process parameters to improve the crystal quality or to reduce production costs, for example. The simplest way is a "try-and-error" approach: use one's intuition to introduce changes in the process specification, perform simulation and evaluate results. A more systematic way is to state an *inverse* problem by indicating 1) what geometry characteristics or process parameters (control parameters) could be varied with valid ranges and 2) what criteria should be used to measure the success of the optimization. The difficulty of the solution of inverse problems is their ill-posedness (Tihonov and Arsenin, 1977). Simulation engine of IEE allows student to solve both direct and inverse problems.

As an example the crystal shape along with the isotherms is shown in Figure 5 after 10 hours since the start of the growth. Note that two regions of parasitic polycrystal grown near the crystal seed and in the crucible corner merge at the later time. Evolution of the porous source during the growth process is illustrated by Figure 6 where distribution of porosity along with streamlines is shown for two time instants: after 2 hours (left) and 10 hours (right). It can be seen that at early time source sublimation is the most intensive near the crucible wall while later dense regions due to recrystallization are formed at the crucible bottom and at the source surface reducing the source efficiency. To increase the powder source utilization and to attain a stable long-term growth one had to search for conditions that provide a uniform temperature distribution across the source, i.e. to solve an inverse problem.

Results of the crucible shape optimization aiming at the minimization of the temperature gradients in the porous source both along the radius and along the axis are presented in Figure 7 and 8. In the first one isotherms in the initial (a) and modified (b,c) crucibles are shown. Temperature distributions in Figure 8 plotted in black, blue and red correspond to cases (a), (b) and (c), respectively.

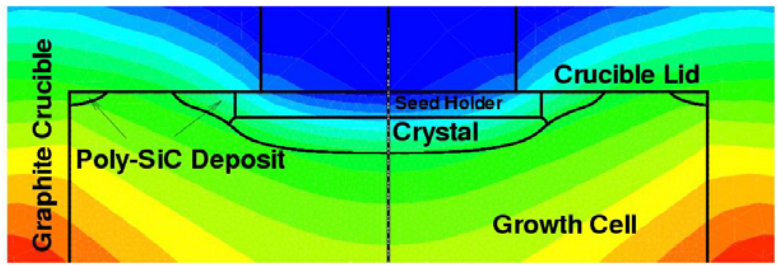

Figure 5: Growth fronts and isotherms after 10 hours

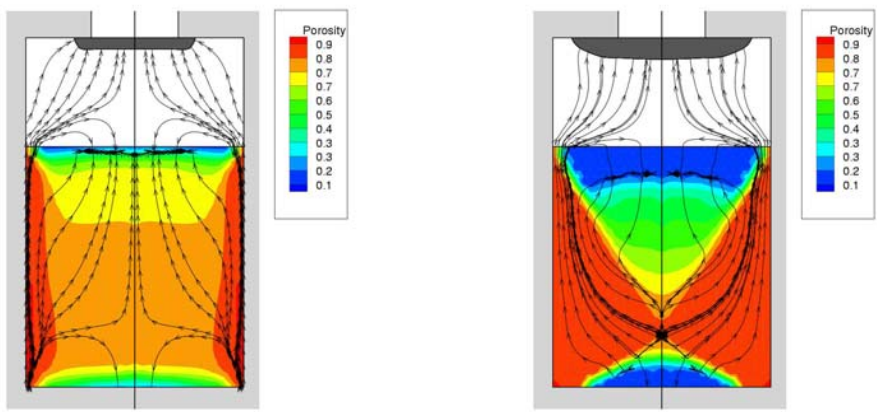

Figure 6: Streamlines and evolution of the source porosity

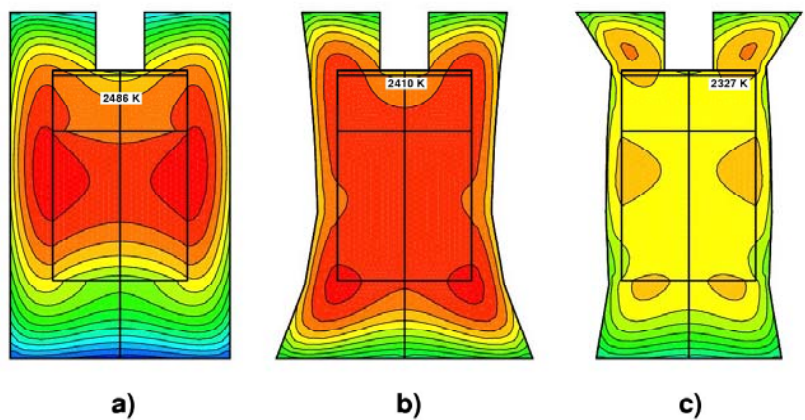

Figure 7: Optimisation of the crucible shape

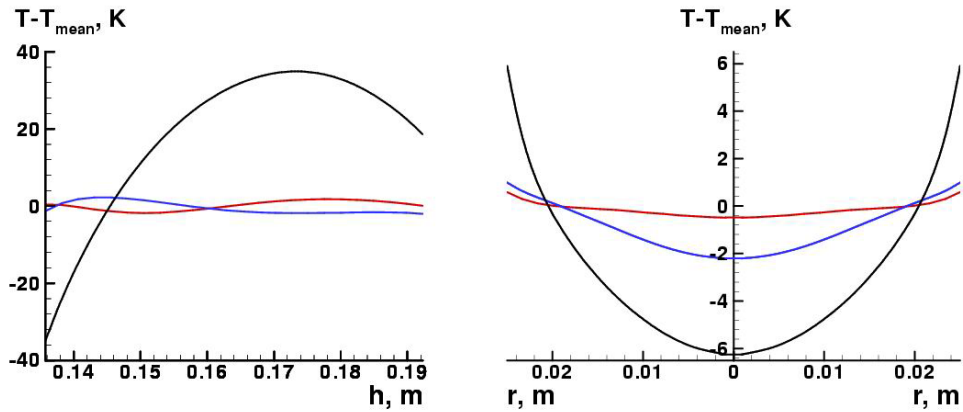

Figure 8: Temperature distribution in the porous source along axis (left) and radius (right)

# 5   Conclusions & Future Work

Problems of modern engineering education are reviewed. A novel tool for Computer-Based Engineering Education (CBEE) – an Integrated Educational Environment (IEE) - based on combination of hypermedia textbook and simulation code is proposed. As a-proof-of-concept IEE "Heat and Mass Transfer in Advanced Semiconductor Technology" has been developed using the software tool Virtual Reactor (ViR) as the simulation engine. The code provides students and engineers with exhaustive information about numerous physical processes responsible for the growth of bulk crystal and its quality. The software simulates global heat transfer in the whole system, diffusion and convective mass transport in the growth cell and porous charge, accounting for crystal faceting, deposit formation and for evolution of the source during long-term growth. Virtual characterization of the growing crystal at various stages of growth provided by the ViR includes analysis of thermo-elastic stress and tracing of threading dislocations with subsequent virtual mapping.

The described IEE is widely used in educational process by Russian technical universities, including Saint-Petersburg State Polytechnic University, Moscow Engineering Physics Institute, Moscow State Technical University. . IEE for advanced electronics and optoelectronics is being developed now. It is planned that first version will include simulation tools for novel devices based on group III-nitride heterostructures: High Electron-Mobility field-effect Transistors (HEMT), Light Emitting Diodes (LED) and Laser Diodes (LD).

**Acknowledgements:** The authors are grateful to Yu.E. Gorbachov for useful discussions. The work has been supported by The Foundation for Assistance to Small Innovative Enterprises (Russia), grants 2072p /4225 (2003) and 3955/p/6183 (2005).

# 7   Footnotes

[1] Korea Ministry of Education (2000). Adapting Education to the Information Age. A White Paper.

[2] Making a European Area of Lifelong Learning a Reality.
http://europa.eu.int/comm/education/policies/lll/life/index_en.html